\documentclass[twocolumn,preprintnumbers,superscriptaddress,unsortedaddress]{revtex4}
\usepackage{amssymb}
\usepackage{amsmath}
\usepackage{graphicx}
\usepackage{dcolumn}
\usepackage{bm}
\usepackage{color,soul}
\usepackage{amsfonts}
\usepackage{mathtools}
\usepackage[thinc]{esdiff}
\usepackage{units}
\usepackage{xcolor}

\newcommand{\func}[1]{\operatorname{#1}}

\begin{document}

\title{Thermal instability in a ferrimagnetic resonator strongly coupled to
a loop-gap microwave cavity}
\author{Cijy Mathai}
\affiliation{Andrew and Erna Viterbi Department of Electrical Engineering, Technion,
Haifa 32000 Israel}
\author{Oleg Shtempluck}
\affiliation{Andrew and Erna Viterbi Department of Electrical Engineering, Technion,
Haifa 32000 Israel}
\author{Eyal Buks}
\affiliation{Andrew and Erna Viterbi Department of Electrical Engineering, Technion,
Haifa 32000 Israel}
\date{\today }

\begin{abstract}
We study nonlinear response of a ferrimagnetic sphere resonator (FSR)
strongly coupled to a microwave loop gap resonator (LGR). The measured
response in the regime of weak nonlinearity allows the extraction of the FSR
Kerr coefficient and its cubic damping rate. We find that there is a certain
range of driving parameters in which the system exhibits instability. In
that range, self-sustained modulation of the reflected power off the system
is generated. The instability is attributed to absorption-induced heating of
the FSR above its Curie temperature.
\end{abstract}

\maketitle

\section{Introduction}

Ferromagnetic and ferrimagnetic resonators \cite{Hill_S227,Lecraw_1311,
Kumar_435802} are widely employed in a variety of microwave (MW) devices,
including narrow band oscillators \cite{Ryte_434}, filters \cite{Tsai_3568},
and parametric amplifiers \cite{Kotzebue_773}. These resonators exhibit a
variety of intriguing physical effects \cite{rezende2020fundamentals},
including Bose-Einstein condensation \cite{Demokritov_430} and
magneto-optical coupling \cite{Zhang_123605, Osada_223601,
Stancil_Spin,Kajiwara_262}. Here we study a strongly coupled hybrid system
composed of a loop gap resonator (LGR) integrated with a ferrimagnetic
sphere resonator (FSR) made of yttrium iron garnet (YIG) \cite%
{Cherepanov_81,Serga_264002}. We focus on the regime of nonlinear response.
In section III below we explore the effect on nonlinear damping in the
region of relatively weak microwave driving. An instability, which is
observed with a much stronger driving, is reported in section IV below, and
a theoretical model, which attributes the instability to a driving-induced
heating, is presented.

Many nonlinear dynamical effects have been observed before in FSRs,
including auto-oscillations \cite{Rezende_1127,Rezende_893}, optical cooling 
\cite{Sharma_087205}, frequency mixing \cite{Jepsen_2627,Morgenthaler_S157}
and bistability \cite{Wang_057202,Wang_224410,Hyde_174423,Suhl_209,Wiese_119}%
. The Suhl instability (of both first and second orders) has been observed
with transverse microwave driving, whereas parallel pumping instability has
been observed with longitudinal driving \cite{cottam2016instability}.
Applications of nonlinearity for quantum data processing have been explored
in \cite{Elyasi_1910_11130,Zhang_023021,Zhang_156401, Tabuchi_083603,
Lachance_070101,Lachance_1910_09096,Tabuchi_729,Kusminskiy_1911_11104}.

\begin{figure}[tbp]
\begin{center}
\includegraphics[width=0.5\textwidth,height=0.5				%
\textheight,keepaspectratio]{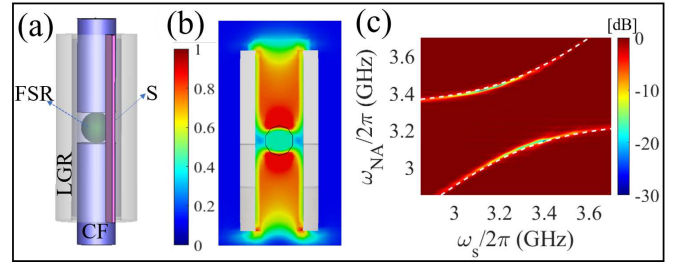}
\end{center}
\caption{{}FSR-LGR coupling: (a) A sketch of the FSR made of YIG having
radius of $R_{\mathrm{s}}=1\func{mm}$ that is integrated inside the aluminum
cylindrical LGR having gap width of $0.3\func{mm}$. The sphere is held by
ceramic ferrules (CFs). A sapphire wafer (labeled as S) is inserted into the
gap to increase the capacitance. (b) The numerically calculated magnetic
field energy density distribution (normalized with respect to the maximum
value) corresponding to driving at the resonance frequency $\protect\omega %
_{ \mathrm{e}}/\left( 2\protect\pi \right) =3.3\func{GHz}$. (c) A VNA
reflectivity $\left\vert S_{11}\right\vert ^{2}$ measurement as a function
of magnon frequency $\protect\omega _{s}$ (proportional to the externally
applied magnetic field). The coupling coefficient $g_{\mathrm{eff}}$ is
extracted from the theoretical fit (white dashed lines) following Eq. (\protect\ref{omega pm}).}
\label{FigLGR}
\end{figure}

Heating a YIG sphere from room temperature to $400\unit{K}$ by microwave
driving having power of $450\unit{mW}$ has been reported in \cite%
{Denton_S300}. At a Curie temperature given by $T_{\mathrm{c}}=560~\unit{K}$%
, YIG undergoes a phase transition between an ordered ferrimagnetic state
(FS) and a disordered paramagnetic state (PS). Thermal instability was
observed in a cavity magneto-mechanical system \cite{Zhang_e1501286}.
Microwave oscillations induced by injecting spin-polarized current \cite%
{Slonczewski_L1} into a magnetic-multilayer structure have been reported in 
\cite{Kiselev_380}. Self-excited oscillations induced by ohmic heating in a $%
\mathrm{Y}_{3}\mathrm{Fe}_{5}\mathrm{O}_{12}$/$\mathrm{Pt}$ bilayer nanowire
have been investigated in \cite{Safranski_1}. Imaging of heating induced by
the spin Peltier effect has been demonstrated in \cite{Daimon_1}.

\section{Loop gap resonator}

With relatively low input power, the main mechanisms responsible for FSR
nonlinear response are magnetic anisotropy \cite{Zhang_987511} and exchange
interaction\cite{Cherepanov_81}. Consider a MW cavity mode having angular
frequency $\omega _{\mathrm{e}}$ and an integrated FSR having radius $R_{%
\mathrm{s}}$. It is assumed that the applied static magnetic field $\mathbf{H%
}_{\mathrm{s}}$ is parallel to the easy axis. In the Holstein-Primakoff
approximation \cite{Holstein_1098} (which assumes that magnetization is
nearly saturated), the Hamiltonian of the system $\mathcal{H}_{\mathrm{D}}$
is expressed as \cite{Wang_224410,Mathai_67001}%
\begin{align}
\hbar ^{-1}\mathcal{H}_{\mathrm{D}}& =\omega _{\mathrm{e}}N_{\mathrm{e}%
}+\omega _{\mathrm{s}}N_{\mathrm{s}}+K_{\mathrm{M}}N_{\mathrm{s}}^{2}  \notag
\\
& +g_{\mathrm{eff}}\left( A_{\mathrm{e}}^{\dag }A_{\mathrm{s}}+A_{\mathrm{e}%
}A_{\mathrm{s}}^{\dag }\right) \ ,  \notag \\
&  \label{H_D}
\end{align}%
where $N_{\mathrm{e}}=A_{\mathrm{e}}^{\dag }A_{\mathrm{e}}$ ($N_{\mathrm{s}%
}=A_{\mathrm{s}}^{\dag }A_{\mathrm{s}}$) is a cavity mode (FSR Kittel mode)
number operator, $\omega _{\mathrm{s}}=\gamma _{\mathrm{g}}H_{\mathrm{s}}$
is the Kittel mode angular frequency, $\gamma _{\mathrm{g}}/2\pi =27.98~%
\unit{GHz}~\unit{T}^{-1}$\ is the gyromagnetic ratio, $K_{\mathrm{M}}=\hbar
\gamma _{\mathrm{g}}^{2}K_{\mathrm{c1}}/\left( V_{\mathrm{s}}M_{\mathrm{s}%
}^{2}\right) $ is the anisotropy-induced Kerr frequency, $K_{\mathrm{c1}}$
is the first-order anisotropy constant, $V_{\mathrm{s}}=4\pi R_{\mathrm{s}%
}^{3}/3$ is the volume of the sphere, $M_{\mathrm{s}}$ is the saturation
magnetization, and $g_{\mathrm{eff}}$ is the cavity-FSR coupling
coefficient. For YIG at room temperature, $M_{\mathrm{s}}=140~\unit{kA}/%
\unit{m}$ and $K_{\mathrm{c1}}=-610~\unit{J}/\unit{m}^{3}$, hence $K_{%
\mathrm{M}}=-2.\,4\times 10^{-8}~\unit{Hz}\times \left( R_{\mathrm{s}%
}/\left( 100\unit{%
\mu%
m}\right) \right) ^{-3}$.

\begin{figure*}[tbp]
\begin{center}
\includegraphics[width=0.75\textwidth,height=0.7								%
\textheight,keepaspectratio]{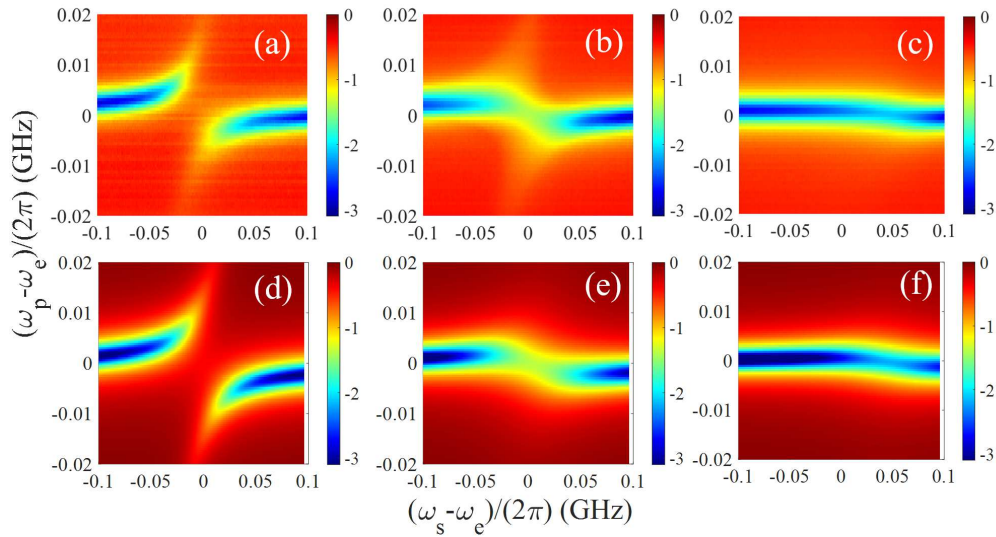}
\end{center}
\caption{Reflection coefficient $\left\vert S_{11}\right\vert ^{2}$ in dB
units for three values of MW input power $P_{\mathrm{p}}$. Panels (a), (b),
and (c) present the experimental data corresponding to MW input powers $P_{ 
\mathrm{p}}$ of -20 dBm, -5 dBm, and +10 dBm, respectively. The second row
[panels (c), (d), and (e)] shows the corresponding theoretical fits that are
obtained from Eq.~ (\protect\ref{cubic SS}). The theoretical fit parameters
are $\protect\gamma _{2\mathrm{e}}=1.5~\unit{MHz}$, $\protect\gamma _{ 
\mathrm{e}}=4~\unit{MHz}$, $\protect\gamma _{\mathrm{s}}=1~\unit{MHz}$, $K_{ 
\mathrm{M}}=6.325~\unit{nHz}$, $\protect\delta _{\mathrm{e}}=35\unit{MHz}$ ,
and $\protect\gamma _{3\mathrm{s}}=0.001~\unit{nHz}$. To obtain a proper
fit, $N_{\mathrm{s}}$ and $g_{\mathrm{eff}}$ are taken as variable values
varying as a function of $P_{\mathrm{p}}$. For $P_{\mathrm{p}}$=$-20$ dBm, $%
-5$ dBm, and $10$ dBm, $N_{\mathrm{s}}$ values are taken as $1\times 10^{19}~%
\unit{m}^{-3}$, $5\times 10^{19}~\unit{m}^{-3}$ and $8\times 10^{19}~\unit{m}%
^{-3}$, and $g_{\mathrm{eff}}$ values are taken as $14\unit{ MHz}$ , $14~%
\unit{MHz}$ and $12\unit{MHz}$, respectively.}
\label{FigSMR}
\end{figure*}

In the linear regime, where the Kerr nonlinearity can be disregarded, the
Hamiltonian $\mathcal{H}_{\mathrm{D}}$\ (\ref{H_D}) can be diagonalized. The
angular frequencies $\omega _{\pm }$ of the two hybrid photon-magnon eigen
modes are given by \cite{Sainz_de_los_Terreros_1906}%
\begin{equation}
\omega _{\pm }=\frac{\omega _{\mathrm{e}}+\omega _{\mathrm{s}}}{2}\pm \sqrt{%
\left( \frac{\omega _{\mathrm{e}}-\omega _{\mathrm{s}}}{2}\right) ^{2}+g_{%
\mathrm{eff}}^{2}}\;.  \label{omega pm}
\end{equation}%
Both angular frequencies $\omega _{\pm }$ are positive provided that $g_{%
\mathrm{eff}}<\sqrt{\omega _{\mathrm{s}}\omega _{\mathrm{e}}}$. Note that
the super-radiance Dicke instability occurs in the ultra-strong coupling
region where $g_{\mathrm{eff}}>\sqrt{\omega _{\mathrm{s}}\omega _{\mathrm{e}}%
}$ \cite{Kirton_1800043}. In the rotating wave approximation (RWA) the Kerr
coefficients $K_{\pm }$ of the hybrid modes having angular frequencies $%
\omega _{\pm }$ are given by Eqs. (\ref{K+}) and (\ref{K-}) of appendix A
[see Eq. (\ref{H_D Kpm Ki})].

In the current experiment, we explore the response for a wide range of the
MW input powers $P_{\mathrm{p}}$. We find that the response is well
described by the Hamiltonian $\mathcal{H}_{\mathrm{D}}$ provided that $P_{%
\mathrm{p}}$ is sufficiently small. However, with sufficiently high $P_{%
\mathrm{p}}$, the FSR temperature $T$ may exceed the Curie temperature $T_{%
\mathrm{c}}$ due to MW absorption-induced heating. We study the response of
the FSR-LGR system to an injected monochromatic pump tone having a frequency
close to resonance. The off reflected power is measured using a spectrum
analyzer (SA). We find that there is a certain zone in the pump frequency -
pump amplitude plane, in which the resonator exhibits limit-cycle (LC)
response resulting in self-sustained modulation of the reflected power. The
observed LC is attributed to thermal instability (TI) \cite{Jeffrey_016213}.

A MW cavity made of an LGR allows achieving a relatively large coupling
coefficient $g_{\mathrm{eff}}$ \cite{Froncisz_515,Zhang_205003}. The MW LGR
schematically shown in Fig. \ref{FigLGR}(a), is made of a hollow concentric
aluminium tube having an inner and outer radii of $R_{\mathrm{LGR}}=1.7\unit{%
mm}$ and $3\unit{mm}$, respectively, and a height of $H_{\mathrm{LGR}}=12%
\unit{mm}$. A sapphire strip of $260\unit{%
\mu%
m}$ thickness has been inserted into the gap in order to increase its
capacitance, which in turn reduces the frequency $f_{\mathrm{e}}$ of the LGR
fundamental mode [$f_{\mathrm{e}}=\omega _{\mathrm{e}}/\left( 2\pi \right)
=3.3\unit{GHz}$ with sapphire] \cite{krupka_387}. An FSR made of YIG having
radius of $R_{\mathrm{s}}=1\unit{mm}$ is held by two ferrules inside the
LGR. The static magnetic field $\mathbf{H}_{\mathrm{s}}$ is applied
perpendicularly to the LGR axis. The LGR-FSR coupled system has been
encapsulated in a metallic rectangular shield made of aluminum. The cavity
is weakly coupled to a loop antenna (LA).

The numerically calculated magnetic energy density distribution
corresponding to the LGR fundamental mode is shown in Fig. \ref{FigLGR}(b).
The calculated density is homogeneous ($\simeq 95\%$) over the FSR volume,
and it is well confined inside the LGR inner volume. Note that for our
device, the LGR inner volume, which is given by $\pi R_{\mathrm{LGR}}^{2}H_{%
\mathrm{LGR}}$, is 4 orders of magnitude smaller than the volume $\lambda _{%
\mathrm{e}}^{3}$, where $\lambda _{\mathrm{e}}=c/f_{\mathrm{e}}$ is the free
space wavelength corresponding to the LGR frequency $f_{\mathrm{e}}$, and $c$%
\ is the speed of light in vacuum. Consequently, the coupling coefficient $%
g_{\mathrm{eff}}$ can be made much larger than typical values obtained with
the commonly employed rectangular cavities \cite{Zhang_156401}, for which
the mode volume commonly has the same order of magnitude as $\lambda _{%
\mathrm{e}}^{3}$.

Based on Eq.~(2) of Ref. \cite{Zhang_156401}, together with the evaluated
energy density shown in Fig. \ref{FigLGR}(b), the calculated value of the
coupling coefficient is found to be $g_{\mathrm{eff}}=176\unit{MHz}$ for the
LGR fundamental mode of frequency $f_{\mathrm{e}}=3.3\unit{GHz}$.
Alternatively, $g_{\mathrm{eff}}$ can be extracted from measurements of MW
reflection coefficient $\left\vert S_{11}\right\vert ^{2}$ as a function of
the Kittel mode frequency $\omega _{\mathrm{s}}/\left( 2\pi \right) $ and
driving frequency $\omega _{\mathrm{NA}}/\left( 2\pi \right) $. Fitting $%
\left\vert S_{11}\right\vert ^{2}$, which is measured at temperature of $3%
\unit{K}$ using a vector network analyzer (VNA), with Eq. (\ref{omega pm})
[see Fig. \ref{FigLGR}(c)] yields the value $g_{\mathrm{eff}}=200\unit{MHz}$%
, which is pretty much close to the value obtained from simulation. Note
that $g_{\mathrm{eff}}$ is only one order of magnitude smaller than the
threshold value corresponding to the super-radiance Dicke instability \cite%
{Kirton_1800043}.

\section{Kerr coefficient and nonlinear damping}

Cavity driving having amplitude $\Omega_{\mathrm{p}}$ and angular frequency $%
\omega_{\mathrm{p}}$ is taken into account by adding a term given by $%
\hbar\Omega_{\mathrm{p}}\left( A_{\mathrm{e}}^{\dag}e^{-i\omega_{\mathrm{p}%
}t}+A_{\mathrm{e}}e^{-i\omega_{\mathrm{p}}t}\right) $ to the Hamiltonian $%
\mathcal{H}_{\mathrm{D}}$ (\ref{H_D}). Steady state solution of the driven
system was calculated in Ref. \cite{Zhang_987511} for the case where damping
is taken into account to first order only. For that case the solution is
found by solving a cubic equation for the FSR dimensionless energy $E_{%
\mathrm{s}}=\left\langle N_{\mathrm{s}}\right\rangle $ [given by Eq. (36) of 
\cite{Zhang_987511}]. We find, however, that the calculated steady state
yields only a moderate agreement with experimental data. Better agreement
can be obtained by taking into account nonlinear damping to cubic order \cite%
{Yurke_5054}. In this approach the cubic equation for $E_{\mathrm{s}}$
becomes%
\begin{equation}
\left( \delta_{\mathrm{s}}^{\prime2}+\gamma_{\mathrm{s}}^{\prime2}\right) E_{%
\mathrm{s}}=\eta\left\vert \Omega_{\mathrm{p}}\right\vert ^{2}\;,
\label{cubic SS}
\end{equation}
where $\delta_{\mathrm{s}}^{\prime}=\delta_{\mathrm{s}}-\eta\delta _{\mathrm{%
e}}+2K_{\mathrm{M}}E_{\mathrm{s}}$, $\delta_{\mathrm{s}}=\omega_{\mathrm{s}%
}-\omega_{\mathrm{p}}$ and $\delta_{\mathrm{e}}=\omega_{\mathrm{e}}-\omega_{%
\mathrm{p}}$\ are driving detuning angular frequencies, $\eta=g_{\mathrm{eff}%
}^{2}/\left( \delta_{\mathrm{e}}^{2}+\gamma_{\mathrm{e}}^{2}\right) $, $%
\gamma_{\mathrm{e}}=\gamma_{1\mathrm{e}}+\gamma_{2\mathrm{e}}$ with $%
\gamma_{1\mathrm{e}}$ ($\gamma_{2\mathrm{e}}$) being the external
(intrinsic) cavity damping rate, $\gamma_{\mathrm{s}}^{\prime}=\gamma_{%
\mathrm{s}}+\eta\gamma_{\mathrm{e}}+\gamma_{3\mathrm{s}}E_{\mathrm{s}}$, $%
\gamma_{\mathrm{s}}$ is the FSR linear damping rate and $\gamma_{3\mathrm{s}%
} $ is the FSR cubic nonlinear damping coefficient. Note that $\left\vert
\Omega_{\mathrm{p}}\right\vert ^{2}$ is proportional to the driving power $%
P_{\mathrm{p}}$ injected into the LA. Note also that when nonlinear damping
is disregarded (i.e. when $\gamma_{3\mathrm{s}}=0$) Eq.~(\ref{cubic SS})
becomes identical to Eq. (36) of \cite{Zhang_987511}.

VNA measurements of the reflection coefficient $\left\vert S_{11}\right\vert
^{2}$ for three different values of $P_{\mathrm{p}}$ are shown in Fig. \ref%
{FigSMR}(a-c). For the data presented in both Fig. \ref{FigSMR} and Fig. \ref%
{FigTI}, the radius of the FSR is $R_{\mathrm{s}}=0.1\unit{mm}$. The
theoretical fit shown in Fig. \ref{FigSMR}(d-f) is based on the cubic
equation (\ref{cubic SS}), which allows the calculation of the dimensionless
energy $E_{\mathrm{s}}$, and on Eq. (3) of Ref. \cite{Zhang_156401}, which
evaluates the reflection coefficient $\left\vert S_{11}\right\vert ^{2}$ as
a function of $E_{\mathrm{s}}$. The values of parameters assumed for the
calculations are listed in the caption of Fig. \ref{FigSMR}. Note the
driving-induced blue shift observed in the magnetic resonance frequency [see
Fig. \ref{FigSMR}(a-c)]. This shift cannot be accurately reproduced
theoretically when nonlinear damping is disregarded.

\section{Thermal instability}

\begin{figure}[tbp]
\begin{center}
\includegraphics[width=3.2in,keepaspectratio]{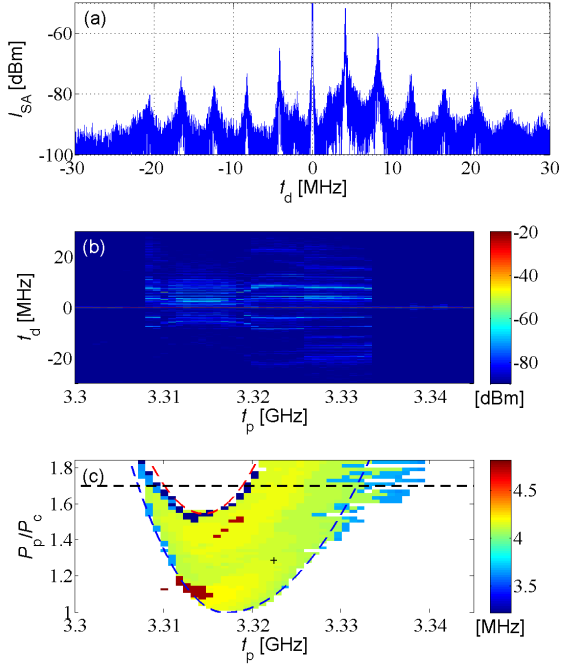}
\end{center}
\caption{{}Thermal instability. (a) Spectral density $I_{\mathrm{SA}}$ of
the signal reflected off the LA, as a function of the detuning frequency $f_{%
\mathrm{d}}$, for the driving frequency $f_{\mathrm{p}}=3.2224~\unit{GHz}$
and normalized driving power $P_{\mathrm{p}}/P_{\mathrm{c}}=1.288$ specified
by the black cross overlaid in (c). (b) Spectral density $I_{\mathrm{SA}}$
in dB as a function of the driving frequency $f_{\mathrm{p}}$ and detuning
frequency $f_{\mathrm{d}}$\ for $P_{\mathrm{p}}/P_{\mathrm{c}}=1.7$
[indicated by the overlaid horizontal dashed line in (c)]. (c) The SB
spacing frequency $\protect\omega _{\mathrm{SM}}/\left( 2\protect\pi \right) 
$ in $\unit{MHz}$ as a function of driving frequency $f_{\mathrm{p}}$ and
normalized driving power $P_{\mathrm{p}}/P_{\mathrm{c}}$. The overlaid blue
(red) dashed line represents the threshold condition $E_{\mathrm{F}}=E_{%
\mathrm{cF}}$ ($E_{\mathrm{P}}=E_{\mathrm{cP}}$). The following values are
assumed for the calculations $\protect\omega _{+\mathrm{F}}/2\protect\pi %
=3.317\unit{GHz}$, $\protect\omega _{+\mathrm{P}}/2\protect\pi =3.314\unit{%
GHz}$, $\protect\gamma _{+\mathrm{F}}=1.3\times \protect\gamma _{+\mathrm{P}%
} $, $\protect\sigma _{\mathrm{F}}/w_{\mathrm{TF}}=2.6\times \protect\sigma %
_{\mathrm{P}}/w_{\mathrm{TP}}$, $\left( K_{+\mathrm{F}}/\protect\gamma _{+%
\mathrm{F}}\right) \left( w_{\mathrm{TF}}/\protect\sigma _{ \mathrm{F}%
}\right) =0.5$ and $K_{+\mathrm{P}}=0$.}
\label{FigTI}
\end{figure}

Further insight can be gained by measuring the spectral density $I_{\mathrm{%
SA}}$\ of the signal reflected off the LA using a SA (see Fig. \ref{FigTI}).
We find that for $P_{\mathrm{p}}>P_{\mathrm{c}}=42.5$ dBm, and for
sufficiently small detuning from resonance, the measured spectral density $%
I_{\mathrm{SA}}$ contains equally-spaced side-bands (SB) on both sides of
the driving frequency $f_{\mathrm{p}}=\omega _{\mathrm{p}}/\left( 2\pi
\right) $ [see Fig. \ref{FigTI}(a)]. We measure the SB spacing frequency $%
\omega _{\mathrm{SM}}/\left( 2\pi \right) $ as a function of the driving
frequency $f_{\mathrm{p}}$ and driving power $P_{\mathrm{p}}$ [see Fig. \ref%
{FigTI}(c)].

The observed equally spaced SBs are attributed to a thermal instability
mechanism that is discussed in Ref.~\cite{Jeffrey_016213}. The phase
transition occurring at the Curie temperature $T_{\mathrm{c}}$ between the
FS and the PS gives rise to a sharp change in the resonance modes of the
hybrid cavity-FSR system. Consider the case where the frequency of the
externally applied driving is tuned very close to the frequency of one the
hybrid system modes. With sufficiently high driving amplitude the
temperature $T$ of the FSR may exceeds the Curie temperature $T_{\mathrm{c}}$
due to driving-induced heating. For that case no steady state with $T<T_{%
\mathrm{c}}$ (i.e. FS) exists. The transition from the FS to the PS
occurring at $T_{\mathrm{c}}$ is expected to give rise to a resonance
frequency shift. Consequently the driving-induced heating is expected to
abruptly drop down, since above $T_{\mathrm{c}}$ the frequency detuning
between the continuous wave external driving and the resonance frequency
becomes larger (in absolute value). Consider the case where the reduced
heating gives rise to a temperature drop below $T<T_{\mathrm{c}}$. For this
case, a steady state with $T>T_{\mathrm{c}}$ (i.e. PS) also becomes
impossible. In the region where no steady state is possible, the temperature
is expected to oscillate around $T_{\mathrm{c}}$. The frequency of
temperature oscillation can be determined from the spacing between the
measured SBs.

For the measurements presented in Fig. \ref{FigTI}, the driving angular
frequency $\omega _{\mathrm{p}}$ is tuned close to $\omega _{+}$. The
analysis is greatly simplified by disregarding the other hybrid eigen mode
having angular frequency $\omega _{-}$. This approximation is applicable in
the strong coupling regime, for which the resonances having angular
frequencies $\omega _{\pm }$ do not overlap [see Eq. (\ref{omega pm})]. In
this approach the FSR-cavity system is treated as a single mode having
angular frequency $\omega _{+}$ $=2\pi \times 3.32~\unit{GHz}$, and Kerr
coefficient $K_{+}=K_{\mathrm{M}}\sin ^{4}\left( \theta _{\mathrm{g}%
}/2\right) $ [see Eq. (\ref{K+})]. The mode damping rate $\gamma _{+}$ $=30~%
\unit{MHz}$ is expressed as $\gamma _{+}=\gamma _{1+}+\gamma _{2+}$, where $%
\gamma _{1+}$ is the coupling coefficient between the driven mode and the
LA, and $\gamma _{2+}$ is the mode intrinsic damping rate (note that $\gamma
_{1+}=\gamma _{2+}$ for critical coupling).

\begin{figure}[tbp]
\begin{center}
\includegraphics[width=3.2in,keepaspectratio]{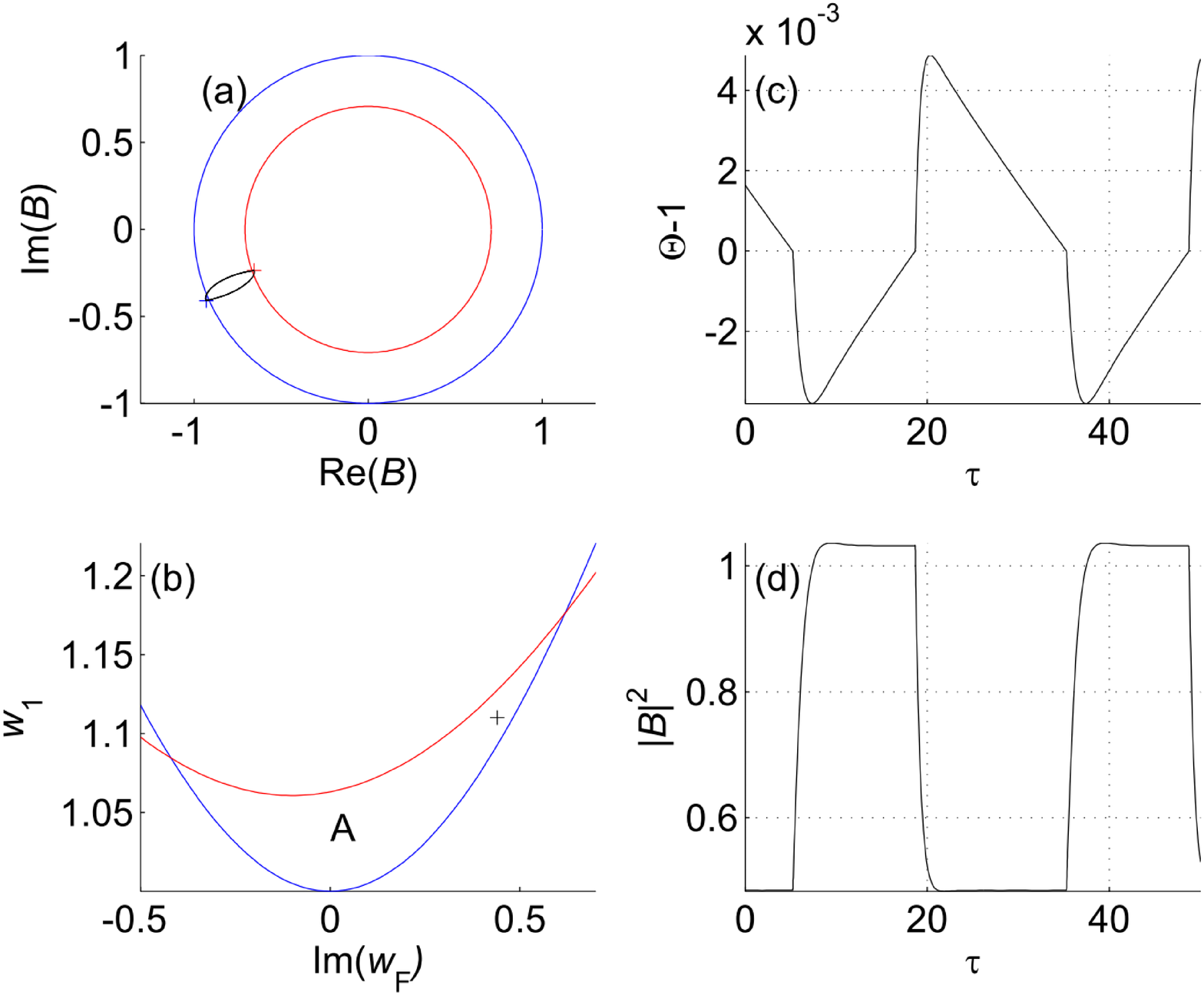}
\end{center}
\caption{{}Limit cycle. (a) Numerical integration of the equations of motion
(\protect\ref{B dot}) and (\protect\ref{Theta dot}) is performed with the
following parameters $\func{Im}\left( w_{\mathrm{F}}-w_{\mathrm{P}}\right)
=-0.1$, $\func{Re}\left( w_{\mathrm{F}}\right) =-1$, $\func{Re}\left( w_{%
\mathrm{P} }\right) =-1.5$, $\protect\sigma _{\mathrm{F}}=0.01$, $\protect%
\sigma _{\mathrm{P}}=0.02$, and $w_{\mathrm{TF}}=w_{\mathrm{TP}}=0.01$. The
values of driving detuning frequency $\func{Im}\left( w_{\mathrm{F}}\right) $
and driving amplitude $w_{1}=w_{1\mathrm{F}}=w_{1\mathrm{P}}$ are indicated
by the black cross in (b). The LC is shown in (a) as a closed curve in the
complex $B$ plane, in (c) as a periodic function of $\Theta -1$ vs. the
normalized time $\protect\tau $, and in (d) as a periodic function $%
\left\vert B\right\vert^{2}$ vs. $\protect\tau $. The plane of driving
frequency and driving amplitude is shown in (b). No steady state solution
exists in the region between the blue and red curves (labeled as A).}
\label{FigTIdA}
\end{figure}

To account for the observed SB, we consider the effect of driving-induced
heating on the FSR magnetic ordering. The externally applied driving gives
rise to a heating power $Q$ given by $Q=2\hslash \omega _{+}\gamma
_{2+}\left\vert B\right\vert ^{2}$, where $B$ is the complex amplitude of
the driven mode (note that nonlinear damping is disregarded here). It is
assumed that the FSR temperature $T$ is uniform, and that the cooling power
due to the coupling between the FSR and its environment at a base
temperature of $T_{0}$ is given by $H\left( T-T_{0}\right) $, where $H$ is
the heat transfer coefficient. The thermal heat capacity of the FSR is
denoted by $C$. It is assumed that all the parameters characterizing the
mode abruptly change at a critical temperature given by $T_{\mathrm{c}}$. In
the adiabatic (diabatic) region, the mode linear damping rate $\gamma _{+}$
is much smaller (larger) than the thermal decay rate $H/C$.

In dimensionless form, system's time evolution is governed by \cite%
{Jeffrey_016213}%
\begin{align}
\dot{B}& =wB-w_{1}\;,  \label{B dot} \\
\dot{\Theta}& =\sigma \left\vert B\right\vert ^{2}-w_{\mathrm{T}}\Theta \;.
\label{Theta dot}
\end{align}%
Overdot denotes a derivative with respect to a dimensionless time $\tau $,
which is related to the time $t$ by $\tau =\gamma _{0}t$, where $\gamma _{0}$
is a constant rate. The dimensionless complex frequency $w$ is given by $%
w=\left( i\left( \omega _{\mathrm{p}}-\omega _{+}-K_{+}\left\vert
B\right\vert ^{2}\right) -\gamma _{+}\right) /\gamma _{0}$, the
dimensionless driving amplitude $w_{1}$ is given by $w_{1}=i\gamma _{0}^{-1}%
\sqrt{2\gamma _{1+}}\Omega _{\mathrm{p}}$, the dimensionless temperature $%
\Theta $ is given by $\Theta =\left( T-T_{0}\right) /\left( T_{\mathrm{c}%
}-T_{0}\right) $, the dimensionless heating coefficient $\sigma $ is given
by $\sigma =2\hslash \omega _{+}\gamma _{2+}\gamma _{0}^{-1}C^{-1}\left( T_{%
\mathrm{c}}-T_{0}\right) ^{-1}$, and the dimensionless thermal rate $w_{%
\mathrm{T}}$ is given by $w_{\mathrm{T}}=\left( H/C\right) /\gamma _{0}$.

The normalized parameters $w$, $w_{1}$, $\sigma $ and $w_{\mathrm{T}}$ are
assumed to have a step function dependence on the temperature. Below (above
) the critical temperature $T_{\mathrm{c}}$, i.e. for $\Theta <1$ ($\Theta
>1 $), they take the values $w_{\mathrm{F}}$, $w_{1\mathrm{F}}$, $\sigma _{%
\mathrm{F}}$ and $w_{\mathrm{TF}}$ ($w_{\mathrm{P}}$, $w_{1\mathrm{P}}$, $%
\sigma _{\mathrm{P}}$ and $w_{\mathrm{TP}}$), respectively. A steady state
(i.e. time independent) solution below (above) the critical temperature $T_{%
\mathrm{c}}$, i.e. in the region $\Theta <1$ ($\Theta >1$), is possible
provided that $E_{\mathrm{F}}<E_{\mathrm{cF}}$ ($E_{\mathrm{P}}>E_{\mathrm{cP%
}}$), where $E_{\mathrm{F}}=\left\vert w_{1\mathrm{F}}/w_{\mathrm{F}%
}\right\vert ^{2}$ and $E_{\mathrm{cF}}=w_{\mathrm{TF}}/\sigma _{\mathrm{F}}$%
\ ($E_{\mathrm{P}}=\left\vert w_{1\mathrm{P}}/w_{\mathrm{P}}\right\vert ^{2}$
and $E_{\mathrm{cP}}=w_{\mathrm{TP}}/\sigma _{\mathrm{P}}$) [see Eqs. (\ref%
{B dot}) and (\ref{Theta dot}) and Fig. \ref{FigTIdA}(b)]. Note that both $%
E_{\mathrm{F}}$ and $E_{\mathrm{P}}$ represent steady state values of Eq. (%
\ref{B dot}) for $\left\vert B\right\vert ^{2}$, whereas both $E_{\mathrm{cF}%
}$ and $E_{\mathrm{cP}}$ represent values of $\left\vert B\right\vert ^{2} $%
, for which $\Theta =1$ is a steady state value of Eq. (\ref{Theta dot}).

Heat can be removed from the FSR by radiation, exchange with the surrounding
air, and exchange with the supporting ferrules, which hold the FSR inside
the LGR. The contributions to the total heat transfer coefficient $H$ due to
radiation, air and the ferrules are denoted by $h_{\mathrm{rad}}S_{\mathrm{s}%
}$, $h_{\mathrm{air}}S_{\mathrm{s}}$ and $H_{\mathrm{fer}}$, respectively,
where $S_{\mathrm{s}}=4\pi R_{\mathrm{s}}^{2}$ is the FSR surface area. The
coefficient $h_{\mathrm{rad}}$ is roughly given by $h_{\mathrm{rad}}\simeq
\alpha _{\mathrm{YIG}}\sigma _{\mathrm{SB}}\left( T_{\mathrm{c}%
}^{4}-T_{0}^{4}\right) /\left( T_{\mathrm{c}}-T_{0}\right) $, where $\alpha
_{\mathrm{YIG}}$\ is the averaged FSR absorption coefficient in the spectral
band corresponding to room temperature $T_{0}\simeq 300\unit{K}$ radiation
(wavelength $\lambda \simeq 10\unit{%
\mu%
m}$), $\sigma _{\mathrm{SB}}=\pi ^{2}k_{\mathrm{B}}^{4}/\left( 60\hbar
^{3}c^{2}\right) $ is the Stefan-Boltzmann constant, $k_{\mathrm{B}}$\ is
the Boltzmann's constant, $\hbar $ is Plank's constant, and $T_{\mathrm{c}%
}=560\unit{K}$ is the YIG Curie temperature. The absorption coefficient
value $\alpha _{\mathrm{YIG}}\simeq 10^{-1}$ \cite{Wood_1038} yields $h_{%
\mathrm{rad}}\simeq 2\unit{W}\unit{m}^{-2}\unit{K}^{-1}$. For ambient
temperature and pressure $h_{\mathrm{air}}\simeq 15\unit{W}\unit{m}^{-2}%
\unit{K}^{-1}$, hence $\left( h_{\mathrm{rad}}+h_{\mathrm{air}}\right) S_{%
\mathrm{s}}\left( T_{\mathrm{c}}-T_{0}\right) \simeq 0.6\unit{mW}$ for a FSR
having radius $R_{\mathrm{s}}=0.1\unit{mm}$. In the region where SB are
observed the induced heating power applied to the FSR is about 3 orders of
magnitudes larger, hence $H\simeq H_{\mathrm{fer}}$, i.e. both radiation and
air have negligibly small contributions, and thus heat is mainly removed by
the ferrules.

The thermal heat capacity of a FSR having radius $R_{\mathrm{s}}=0.1\unit{mm}
$ and volume $V_{\mathrm{s}}=4\pi R_{\mathrm{s}}^{3}/3$\ is given by $%
C=2.9\times 10^{6}~\unit{J}\unit{K}^{-1}\unit{m}^{-3}\times V_{\mathrm{s}%
}=1.2\times 10^{-5}\unit{J}\unit{K}^{-1}$ \cite{hofmeister_45}, hence the
thermal decay rate is roughly given by $H/C\simeq 320~\unit{Hz}\times \left(
Q_{\mathrm{c}}/\unit{W}\right) \left( \left( T_{\mathrm{c}}-T_{0}\right)
/\left( 260~\unit{K}\right) \right) ^{-1}$, where $Q_{\mathrm{c}}$ is the
heating power applied to the FSR, for which the steady state temperature is $%
T_{\mathrm{c}}$. Hence for the current device $\left( H/C\right) /\gamma
_{+}\simeq 10^{-5}$, and thus the diabatic approximation is applicable.

A typical limit cycle (LC) in the diabatic regime is shown in Fig. \ref%
{FigTIdA}. The LC is calculated by numerically integrating the equations of
motion (\ref{B dot}) and (\ref{Theta dot}). The blue (red) cross shown in
Fig. \ref{FigTIdA}(a) indicates the steady state value $w_{1}/w$ of $B$
corresponding to the FS (PS), i.e. for $\Theta <1$ ($\Theta >1$), and the
blue (red) circle represents the relation $\left\vert B\right\vert ^{2}=E_{%
\mathrm{cF}}$ ($\left\vert B\right\vert ^{2}=E_{\mathrm{cP}}$). In the plane
of driving frequency and driving amplitude, which is shown in Fig. \ref%
{FigTIdA}(b), the blue and red curves are derived from the relations $E_{%
\mathrm{F}}=E_{\mathrm{cF}}$ and $E_{\mathrm{P}}=E_{\mathrm{cP}}$,
respectively. In the region labeled as A, no steady state solution to Eqs. (%
\ref{B dot}) and (\ref{Theta dot}) exists. The LC period time $\tau _{%
\mathrm{LC}}$ can be calculated by integrating Eqs. (\ref{B dot}) and (\ref%
{Theta dot}) over a single period. In the diabatic limit, one finds that $%
\tau _{\mathrm{P}}\simeq \left\vert w_{\mathrm{P}}\right\vert
^{-1}+\left\vert w_{\mathrm{F}}\right\vert ^{-1}$. The measured value of LC
frequency roughly agrees with this theoretical estimation.

\section{Summary}

In summary, we demonstrate that relatively large coupling coefficient $g_{%
\mathrm{eff}}$ can be obtained by employing an LGR having mode volume much
smaller than $\lambda _{\mathrm{e}}^{3}$. The response of the system in the
weak nonlinear regime allows the extraction of the Kerr coefficient $K_{%
\mathrm{M}}$ and the cubic nonlinear damping rate $\gamma _{3\mathrm{s}}$.
An instability is revealed by driving the system with a relatively high
input power. Above the instability threshold the response of the system to
an externally applied monochromatic driving exhibits self-modulation. The
instability, which is attributed to driving-induced heating, occurs in a
region where the response has no steady state value. Further study will be
devoted to developing sensors that exploit this instability for performance
enhancement.

\section{Acknowledgments}

This work was supported by the Israeli science foundation, the Israeli
ministry of science, and by the Technion security research foundation.

\appendix

\section{Rotating wave approximation}

The Hamiltonian (\ref{H_D}) can be expressed as%
\begin{equation}
\hbar ^{-1}\mathcal{H}_{\mathrm{D}}=\left( 
\begin{array}{cc}
A_{\mathrm{e}}^{\dag } & A_{\mathrm{s}}^{\dag }%
\end{array}%
\right) M\left( 
\begin{array}{c}
A_{\mathrm{e}} \\ 
A_{\mathrm{s}}%
\end{array}%
\right) +K_{\mathrm{M}}N_{\mathrm{s}}^{2}\ ,
\end{equation}%
where the $2\times 2$ matrix $M$ is given by%
\begin{equation}
M=\left( 
\begin{array}{cc}
\omega _{\mathrm{e}} & g_{\mathrm{eff}} \\ 
g_{\mathrm{eff}} & \omega _{\mathrm{s}}%
\end{array}%
\right) \ .
\end{equation}%
The eigenvalues $\omega _{\pm }$ of the matrix $M$ are given by $\omega
_{\pm }=\omega _{\mathrm{m}}\pm \sqrt{\omega _{\mathrm{d}}^{2}+g_{\mathrm{eff%
}}^{2}}$ [see Eq. (\ref{omega pm})], where $\omega _{\mathrm{m}}=\left(
\omega _{\mathrm{e}}+\omega _{\mathrm{s}}\right) /2$ and $\omega _{\mathrm{d}%
}=\left( \omega _{\mathrm{e}}-\omega _{\mathrm{s}}\right) /2$. The matrix $M$
can be expressed as%
\begin{equation}
M=\omega _{\mathrm{m}}\left( 
\begin{array}{cc}
1 & 0 \\ 
0 & 1%
\end{array}%
\right) +\sqrt{\omega _{\mathrm{d}}^{2}+g_{\mathrm{eff}}^{2}}\left( 
\begin{array}{cc}
\cos \theta & \sin \theta \\ 
\sin \theta & -\cos \theta%
\end{array}%
\right) \;,
\end{equation}%
where%
\begin{equation}
\tan \theta =\frac{g_{\mathrm{eff}}}{\omega _{\mathrm{d}}}\;.
\label{tan(theta)}
\end{equation}%
The transformation%
\begin{equation}
\left( 
\begin{array}{c}
A_{\mathrm{e}} \\ 
A_{\mathrm{s}}%
\end{array}%
\right) =U\left( 
\begin{array}{c}
A_{+} \\ 
A_{-}%
\end{array}%
\right) \;,
\end{equation}%
where%
\begin{equation}
U=\left( 
\begin{array}{cc}
\cos \frac{\theta }{2} & -\sin \frac{\theta }{2} \\ 
\sin \frac{\theta }{2} & \cos \frac{\theta }{2}%
\end{array}%
\right) \;,  \label{U=}
\end{equation}%
which diagonalizes the linear part of $\mathcal{H}_{\mathrm{D}}$, yields%
\begin{equation}
\hbar ^{-1}\mathcal{H}_{\mathrm{D}}=\omega _{+}N_{+}+\omega _{-}N_{-}+K_{%
\mathrm{M}}\left( A_{\mathrm{s}}^{\dag }A_{\mathrm{s}}\right) ^{2}\ ,
\label{H_D K}
\end{equation}%
where $A_{\mathrm{s}}=A_{+}\sin \left( \theta /2\right) +A_{-}\cos \left(
\theta /2\right) $, and where $N_{\pm }=A_{\pm }^{\dag }A_{\pm }$.

In the rotating wave approximation (RWA) the Hamiltonian (\ref{H_D K})
becomes%
\begin{eqnarray}
\hbar ^{-1}\mathcal{H}_{\mathrm{D}} &=&\omega _{+}N_{+}+\omega _{-}N_{-} 
\notag \\
&&+K_{+}N_{+}^{2}+K_{-}N_{-}^{2}+K_{\mathrm{i}}N_{+}N_{-}\ ,  \notag \\
&&  \label{H_D Kpm Ki}
\end{eqnarray}%
where the Kerr coefficients $K_{\pm }$ are given by 
\begin{eqnarray}
K_{+} &=&K_{\mathrm{M}}\sin ^{4}\frac{\theta }{2}\ ,  \label{K+} \\
K_{-} &=&K_{\mathrm{M}}\cos ^{4}\frac{\theta }{2}\ ,  \label{K-}
\end{eqnarray}
and the inter-mode Kerr coefficient $K_{\mathrm{i}}$ is given by $K_{\mathrm{%
i}}=K_{\mathrm{M}}\sin ^{2}\theta $.

\bibliographystyle{IEEEtran}
\bibliography{Eyal_Bib}

\end{document}